\documentclass[twocolumn,prl,aps]{revtex4}
\usepackage{amssymb,amsmath,amsthm}
\usepackage{graphicx}
\usepackage{verbatim}

\begin{document}
\title{Zig-zag instability of an Ising wall in liquid crystals}
\author{C. Chevallard, M.G. Clerc, P. Coullet\thanks{
Professeur \`a l'Institut Universitaire de France.}, J.-M. Gilli}
\address{Institut Non Lin\'{e}aire de Nice, UMR 6618 CNRS-UNSA,
 1361 Route des
Lucioles, F-06560 Valbonne, France.}

\begin{abstract}
We present a theoretical explanation for the interfacial zigzag instability
that appears in anisotropic systems. Such an instability has been 
experimentally highlighted
for an Ising wall formed in a nematic liquid crystal cell under
homeotropic anchoring conditions. From an envelope equation, relevant
close to the Fr\'eedericksz transition, we have derived an 
asymptotic equation describing the
interface dynamics in the vicinity of its bifurcation.
The asymptotic limit used accounts for a strong difference between
two of the elastic constants. The model is characterized by a
conservative order parameter which satisfies a Cahn-Hilliard
equation. It provides a good qualitative understanding of the
experiments.
\end{abstract}


\maketitle

The zig-zag instability, undergone by straight rolls in
two-dimensional extended systems like Rayleigh-B\'{e}nard
convection \cite{manneville} or electroconvection \cite{kramer}
fluid systems, have been extensively studied over the last
decades. A similar zig-zag instability affecting anisotropic
interfaces has recently aroused considerable interest in different
fields, for example, in gas discharge system \cite{astrov},
crystal growth \cite{nepomnyashchy}, rifts in spreading wax layer
\cite{bodenschatz}, Ising wall in nematic and cholesteric liquid
crystals \cite{Chevallard2000} and chevrons layer structure of
smectic liquid crystals \cite{limat}. In this letter, we present a
theoretical explanation for this interfacial zigzag instability
and show how it justifies the Ising wall dynamics in liquid
crystal samples. Besides, the instability is experimentally
characterized for a thin layer of nematic liquid crystal
homeotropically anchored. From an amplitude equation, which is
valid in the vicinity of the Fr\'{e}edericksz transition, we
derive a solvability equation that describes the interface
dynamics at the onset of the bifurcation. This model is in
qualitative agreement with the experimental results, even though
these latter were obtained far from the Fr\'{e}edericksz
transition. Numerical simulations allow to strengthen the
relevance of the model.

In two-dimensional extended systems, the dynamics of a straight
interface can be entirely characterized by the dynamics of the
interface position $P$, function that parametrizes the interface
in space and time. Henceforth we consider that the variables $x$
and $y$ describe respectively the transversal and longitudinal
directions of the straight interface. As a consequence of the
translational invariance of the interface ($P\longrightarrow
P+P_{o}$) and of the space reflection symmetry in the tangential
direction to the interface ($y\longrightarrow -y$), the order
parameter $P(t,y)$ shows a spatio-temporal evolution which is
modelled by the Kuramoto-Sivashinky equation \cite{kuramoto}, as
long as one assumes that, the interface only depends on the
tangential direction. If one considers moreover the space
reflection symmetry in the transversal direction
($P\longrightarrow -P$ ), then the order parameter satisfies the
following diffusion equation

\begin{equation}
\partial _{t}P=\varepsilon P_{yy}.  \label{E-difusion}
\end{equation}
When $\varepsilon $ is small (positive or negative), the order
parameter dynamics is described by the asymptotic equation

\begin{equation}
\partial _{t}P=\varepsilon P_{yy}+3P_{y}^{2}P_{yy}-P_{yyyy}
\label{E-Interface}
\end{equation}
where $\varepsilon $ is the diffusion (antidiffusion, $\varepsilon
<0$), $ P_{y}^{2}P_{yy}$ the nonlinear diffusion and the last term
the hyperdiffusion.

In the preceding expressions, we have only considered transversal
perturbations to the straight interface and adiabatically
eliminated the diffusive behavior in the longitudinal direction,
approximations that are both correct for anisotropic systems. In
isotropic systems, a straight interface can be linearly unstable
($\varepsilon <1$) along the longitudinal direction as a result of
the interface symmetry. However, in this case, nonlinear terms
appear simultaneously ($\partial _{x}P\partial _{yy}P$, and
$\partial _{y}P\partial _{xy}P$\ evaluated at the interface) that
destabilize the solutions of eq. (\ref{E-Interface}). Due to these
nonlinear terms, the vertices,\ shown on figures \ref{fig2} and
\ref{fig4}, will become fingers that propagate along the
transversal direction \cite {CocoMar}. Therefore all physical
systems that show a zig-zag instability leading to a zigzag
coarsening dynamics are anisotropic \cite
{astrov,nepomnyashchy,bodenschatz,limat}.

In eq. (\ref{E-Interface}) we have assumed that the nonlinear and
fourth-order derivative terms saturate the instability, which
imposed the signs of these two terms. Note that this equation is a
continuity equation that expresses the conservation of the area
enclosed by the curve $P(y,t)$ and the y-axis. This law stands
from the fact that the interface connects two symmetrical states
($P\longrightarrow -P$) and hence, when one part of the interface
moves in one direction, another part moves simultaneously in the
opposite direction. Moreover, this equation is variational,{\it
i.e}. it can be rewritten in the following form

\begin{equation}
\partial _{t}P=-\frac{\delta {\cal F}}{\delta P}\text{, \ \ \ }{\cal F}
[P]=\int dy\left\{ \varepsilon
\frac{P_{y}^{2}}{2}+\frac{P_{y}^{4}}{4}+\frac{
P_{yy}^{2}}{2}\right\}
\end{equation}
where the {\it free energy} ${\cal F}$ only depends on the order
parameter derivatives. Introducing the variable $\Lambda \equiv
P_{y}$, which describes the local tilt of the interface with
respect to the original orientation of the straight interface, the
latter equation is reduced to

\begin{equation}
\partial _{t}\Lambda =\partial _{yy}\left( \varepsilon \Lambda +\Lambda
^{3}-\Lambda _{yy}\right)  \label{Cahn-hilliard}
\end{equation}
This equation is the well-known Cahn-Hilliard equation \cite{cahn}, which
describes the phase separation dynamics in conservative systems. The
dynamical behavior of this equation has been studied by many authors (see
for example \cite{kawasaki,Bates} and the references therein). It is
noteworthy that the relevant conservation law for the dynamics is actually

\begin{equation}
M=\int \Lambda \left( y,t\right) dy
\end{equation}
and not the integral of $P$, as a consequence of the translational
invariance. When $M$ is zero, the dynamical evolution, deduced
from the theoretical analysis, is in good quantitative agreement
with the experiments, as we shall see later. A wave number of
order $\sqrt{ -\varepsilon /2}$ arises initially in the system.
Later on, the system exhibits a coarsening dynamics (see fig.
\ref{fig2}) during which the wavelength increases. New periodic
solutions emerge with a logarithmically slow growth rate
\cite{Bates}.

To find the steady solutions, one has to minimize the free energy
taking into account the area conservation. This can be done by
considering a Lagrange multiplier ($\lambda $). Hence, the free
energy takes the form \cite{Elliot}

\begin{equation}
{\cal G}[\Lambda ]=\int dy\left\{ \varepsilon \frac{\Lambda
^{2}}{2}+\frac{ \Lambda ^{4}}{4}+\frac{\Lambda
_{y}^{2}}{2}+\lambda \Lambda \right\}
\end{equation}
The family of solutions that minimize this free energy satisfies the
differential equation

\begin{equation}
\Lambda _{yy}=\lambda +\varepsilon \Lambda +\Lambda ^{3}
\end{equation}
Obviously such solutions also satisfy eq. (\ref{Cahn-hilliard}).
The global minimum of the free energy ${\cal G}$ is either the
homoclinic solution (which is an interface with three facets) or
the constant solution. This one, as well as the facets, can be any
of the local minima of the potential $ V=\lambda \Lambda
+\varepsilon \frac{\Lambda ^{2}}{2}+\frac{\Lambda ^{4}}{4}$ . We
remark that these solutions have different Lagrange multiplier
values.

Let's consider now an anisotropic nematic liquid crystal\ sample
that exhibits experimentally an interfacial zig-zag instability,
as emphasized below. This instability will be theoretically
described in the framework of the nonlinear elasticity theory of
liquid crystals.

\begin{figure}[tbp]
\centerline{
\includegraphics[scale=0.6,angle=0]{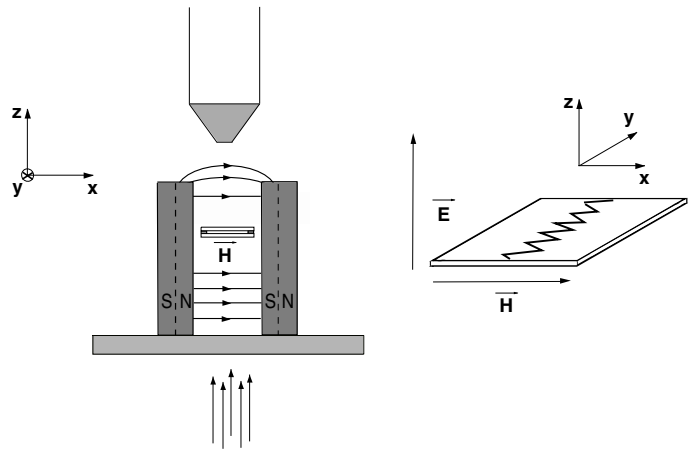}}
\caption{Experimental set-up: the sample of liquid crystal is
placed between two permanent magnets which determine the
homogeneous magnetic field $\vec{H} $. The physical phenomena are
observed through a polarizing microscope. } \label{fig1}
\end{figure}

In the experiments, we used a cyanobiphenyl compound (5CB) whose
anisotropic physical properties at $25^{\circ }C$ are: elastic
constants $ K_{1}=6.3$, $K_{2}=4.1$, $K_{3}=8.4$ ($10^{-7}$
dynes); dielectric anisotropy $\epsilon _{a}=11.3$; diamagnetic
anisotropy $\chi _{a}=1.142$ ($ 10^{-7}$cm$^{3}g^{-1}$);
rotational viscosity $\gamma _{1}\sim 10^{-2}$Pa.s.

The samples are made of two glass plates separated by thin mylar
spacers that determine the cell thickness (between $50$ and
$250~\mu m$). The glass surfaces are treated with lecithin to
provide homeotropic anchoring. The sample is subjected to a
sinusoidal vertical electric field $ \vec{E}=E\vec{e}_{z}$
($V_{eff.}\sim 0-9V$) with a high frequency ($\sim 5kHz $) in
order to avoid charge injection or electroconvection phenomena.
Moreover, two permanent magnets induce a homogeneous horizontal
magnetic field in the sample. The field magnitude can be changed
by bringing the magnets nearer or farther. Its maximum value is
about $0.55$ Tesla.

The application of a horizontal magnetic field can induce, for a
high enough magnitude, a partial reorientation of the bulk
molecules along the field direction. Due to the twofold degeneracy
of the bifurcated state, domains of opposite orientation may be
created \cite{helfrich}. The interface between a pair of domains
is called {\sl Ising wall} \cite {gilli}. An Ising wall in a
splay-bend configuration is formed by using the flux lines
curvature in a region where the field is inhomogeneous (see fig.
\ref{fig1}). Then, it is quenched into the area between the two
magnets, where the field is homogeneous. Its width is determined
by the distance to the threshold of the Fr\'{e}edericksz
transition and can be modified by variations of the fields
magnitude.

\begin{figure}[tbp]
\centerline{
\includegraphics[scale=0.5,angle=0]{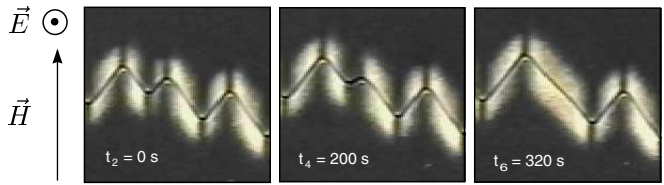}}
\caption{{\protect\small {Spinodal decomposition of the interface
observed through a microscope with crossed polarizers.}}}
\label{fig2}
\end{figure}

The interface behavior is observed through a polarizing microscope
(see fig. \ref{fig1}). Video films or numerical images (see fig.
\ref{fig2}) can be registered thanks to a 3CCD camera placed on
the top of the microscope.

The experiments have been carried out far from the
Fr\'{e}edericksz threshold, since near the transition any little
imperfection on the parallelism between the sample and the
magnetic field plane makes the wall drift towards one edge of the
cell.

The splay-bend Ising wall is ever unstable, for the control
parameters that we used experimentally, when it is thrust into the
homogeneous area of the magnetic field. Initially, the interface
develops an instability characterized by a well-defined
wavelength, which is determined by the experimental parameters.
Later on, the sinusoidal interface becomes an angled line composed
of pieces of wall turned with angles $\pm \Psi _{0}$ (see fig.
\ref{fig2}). Two adjacent pieces, whose orientations are opposite,
are connected by a region of strong curvature of the line that we
called {\sl kink}. The dynamics consists then in reassembling
domains of even orientation, the angle of the ''zig'' and ''zag''
facets staying unchanged (coarsening dynamics). This process
occurs thanks to annihilations of kinks and without characteristic
lengthscale. Actually, the averaged domain size increases
regularly in time (see fig. \ref{fig2}). The dynamics, which tends
to separate the zig and zag facets, is the one-dimensional
counterpart of the spinodal decomposition dynamics observed in
conservative binary mixtures \cite{cahn}.

\begin{figure}[tbp]
\centerline{
\includegraphics[scale=0.4,angle=0]{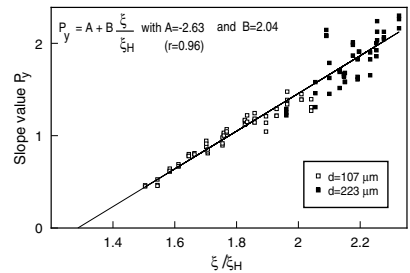}}
\caption{{\protect\small {Experimental curve giving the slope
$P_{y}$ of the interface versus the dimensionless ratio
$\frac{\protect\xi }{\protect\xi _{H}}$ ($\protect\xi $ is the
electro-magnetic coherence length and $\protect \xi _{H}$ the
magnetic coherence length).}}} \label{fig3}
\end{figure}

\begin{figure}[tbp]
\centerline{
\includegraphics[scale=0.5,angle=0]{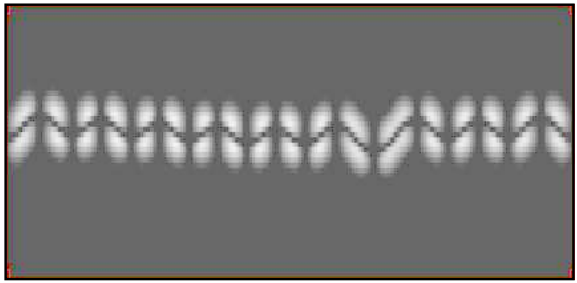}}
\caption{{\protect\small {Numerical simulation of the eq.(8) for a
high anisotropy of elasticity~: $\frac{K_{2}}{K_{1}}=0.056$ ,
$K_{3}=1$ , $a=1$. The picture shows the Y-component of the
director in the system.}}} \label{fig4}
\end{figure}

The instability is triggered by the elastic anisotropy of the
liquid crystal whose influence is emphasized in the wall. Indeed,
the involved distortions depend on the orientation of the
interface with respect to the magnetic field direction. For most
of the usual compounds, the energy cost of a wall aligned with the
field (twist wall) is lower than the one of a wall perpendicular
to the field (splay-bend wall) since $K_{2}<K_{1}$. Consequently,
the system (splay-bend wall) reduces its energy by changing the
direction of the wall. A global rotation of the interface can not
occur in an infinite medium. Therefore, the interface is forced to
rotate locally and is divided into facets turned with opposite
angles $\pm \Psi _{0}$. These angles, whose values are determined
by the control parameters (see fig. \ref{fig3}), result from
several effects. More precisely, the rotation is favored by the
elasticity but is made difficult by the resulting elongation of
the interface along its original direction and by the escape of
the molecules from the vertical plane containing the magnetic
field. The local reorientation generates many defects (kinks),
which will eventually disappear since this leads to a decrease in
energy.

The experiments have been performed far from the Fr\'{e}edericksz
transition. However, for the sake of simplicity, the theoretical
approach has been developed near this transition. At the onset of
the bifurcation, the system is described by the two-dimensional
Landau equation

\begin{equation}
\partial _{T}Z=\varepsilon Z-bZ^{3}+\left( K_{1}\partial
_{x}^{2}+K_{2}\partial _{y}^{2}\right) Z.
\end{equation}
where $\varepsilon \equiv -\epsilon _{a}E^{2}-K_{3}\frac{\pi
^{2}}{d^{2}} +\chi _{a}H^{2}$, $b\equiv \frac{1}{2}\left(
K_{1}-\frac{3}{2}K_{3}\right) \frac{\pi
^{2}}{d^{2}}-\frac{3}{4}\epsilon _{a}E^{2}+\frac{3}{4}\chi
_{a}H^{2}$, $T=t/\gamma _{1}$, and the director takes the form
$\vec{n} =\left( Z\cos (\frac{\pi z}{d}),0,1-\frac{Z^{2}}{2}\cos
^{2}(\frac{\pi z}{d} )\right) $. This equation contains wall-type
solutions. When both diffusion constants ($K_{1}/\gamma _{1}$ and
$ K_{2}/\gamma _{1}$) are of the same order of magnitude, these
solutions are always stable. In order to explain the appearance of
the interfacial instability, one has to consider that $K_{2}$ is
much smaller than the other constants $K_{1}$ and $K_{3}$
($K_{2}\sim \varepsilon $). Then, the solution $Z=0$ is marginal
with respect to perturbations in the $y$-direction. Note that the
interface identifies with the center of the Ising wall, which is a
curve where $Z=0$ (homeotropic state). Hence, higher-order terms
must be considered in order to study the stability of the
interface, and the dynamical equation reads now
\begin{eqnarray}
\partial _{T}Z &=&{\varepsilon Z-bZ^{3}+\left( K_{1}\partial
_{x}^{2}+K_{2}\partial _{y}^{2}\right) Z}+{\frac{K_{1}^{2}}{a}\partial
_{x^{2}y^{2}}Z}  \nonumber \\
&&+{\frac{K_{1}^{3}}{a^{2}}\partial _{x^{2}y^{4}}Z}+{\frac{3}{4}K_{3}}\text{
}\left( {Z\left( \partial _{y}Z\right) ^{2}-}\frac{{Z}^{2}}{2}\partial
_{yy}Z\right) ,  \label{E-formanormal}
\end{eqnarray}
where we have considered the asymptotic limit $Z\sim \varepsilon ^{1/2}$, 
$\partial _{y}^{2}Z\sim \mu \varepsilon ^{1/2}$, $\partial _{x}^{2}Z\sim
\varepsilon ^{3/2}$, $K_{1}\sim K_{3}\sim 1$, and $K_{2}\sim \varepsilon $,
with $\varepsilon \ll \mu \ll 1$, $a\equiv $$\epsilon _{a}E^{2}+K_{3}
\frac{\pi ^{2}}{d^{2}}\hspace{0.1cm}$and $\mu \equiv \left( \frac{K_{2}}{
\varepsilon }-\frac{2K_{1}}{5a}\right)$. Then, the expression of the local
director is given by
\[
\vec{n}=\left(
\begin{array}{c}
n_{x} \\
n_{x} \\
n_{z}
\end{array}
\right) =\left(
\begin{array}{c}
Z\cos \left( \frac{\pi z}{d}\right) \\
\left( {\frac{K_{1}}{a}\partial _{xy}}Z+{\frac{K_{1}^{2}}{a^{2}}\partial
_{xy^{3}}}Z\right) \cos \left( \frac{\pi z}{d}\right) \\
1-\frac{Z^{2}}{2}\cos ^{2}\left( \frac{\pi z}{d}\right)
\end{array}
\right)
\]
In this asymptotic limit, the Ising wall solution reads $Z=\sqrt{
\varepsilon /b}\tanh \left( \sqrt{\varepsilon /2K_{1}}\left(
x-x_{o}\right) \right) $. In order to describe the dynamical
behaviour of the wall under perturbations, we introduce the
following ansatz
\begin{equation}
Z\left( x,P(y,T)\right) \equiv \sqrt{\frac{\varepsilon }{b}}\tanh \left(
\sqrt{\frac{\varepsilon }{2K_{1}}}x-P(y,T)\right) +w(x,P)
\end{equation}
where $w(x,P)$ is a small perturbative{\bf \ }term. One can show
that, $P$ satisfies the following solvability equation

\begin{equation}
\partial _{T}P=D_{1}P_{yy}+D_{2}P_{y}^{2}P_{yy}-D_{3}P_{yyyy}
\label{interface_eq}
\end{equation}
with $D_{1}\equiv \mu \varepsilon =\left(
K_{2}-\frac{2K_{1}\varepsilon }{5a} +\frac{3K_{3}\varepsilon
}{40b}\right) $, $D_{2}\equiv \frac{48K_{1}^{2}}{7}$
$\frac{\varepsilon }{a^{2}}$, $D_{3}\equiv
\frac{2K_{1}^{2}}{5}\frac{ \varepsilon }{a^{2}}$.

Thus, the characteristic time-scale of the interface dynamics
$\left( \mu ^{-1}\varepsilon ^{-1}\right) $ is smaller than the
time-scale of the interface formation $\left( \varepsilon
^{-1}\right) $, which is consistent with experimental
observations. The expression of $D_{1}$ indicates that the elastic
anisotropy can give rise to the instability ($ D_{1}<0$).
Numerical simulations of eq. (\ref{E-formanormal}) confirms the
latter calculations (see fig. \ref{fig4}).

As a consequence of the finite size of liquid crystal sample, one
experimentally observes a global rotation of the interface, which
is superimposed to the coarsening dynamics. This favours one
orientation of the domains, but does not change the angle of the
facets.

In conclusion, we have presented a theoretical explanation of the
interfacial zig-zag instability that appears in anisotropic
systems. The instability of an Ising wall formed in a nematic
liquid crystal has been more specifically investigated, both
experimentally and theoretically. Close to the Fr\'{e}edericksz
transition, this instability is described by the Cahn-Hilliard
equation, and\ leads to the facets separation and coarsening
dynamics. The asymptotic limit used in the derivation of this
model emphasizes the role of the elastic anisotropy. Such a limit
contains all the ingredients required to give an adequate
description of the experimental observations.

\bigskip

The authors acknowledge S. Thiberge for fruitful discussions. One
of us P.C. thanks the support of the ''Institut Universitaire de
France'' and M.C. thanks he support of C\'{a}tedra Presidencial.
The simulation software developed at the laboratory INLN in France
has been used for all the numerical simulations presented in this
paper.


\bigskip

\begin{references}
\bibitem{manneville}  MANNEVILLE P. and PIQUEMAL J.-M., {\sl Phys. Rev. A}
{\it ,} {\bf 28} (1983) 1774; CROSS M. C. and HOHENBERG P. C.,
{\sl Rev. Mod. Phys.}, {\bf 65 }(1993).

\bibitem{kramer}  RIBOTTA R., JOETS A., and LEI L., {\sl Phys. Rev. Lett.}
{\bf 56} (1986) 1595; BODENSCHATZ E., KAISER M, KRAMER L., PESCH W., WEBER
A., ZIMMERMANN W., {\sl New Trends in Nonlinear Dynamics and Pattern-Forming
\ Phenomena}, edited by P. COULLET and P. HUERRE (Plenum Press, New York)
1990.

\bibitem{astrov}  ASTROV Yu. A., AMMELT E. and PURWINS H.G., {\sl Phys. Rev.
Lett.}, {\bf 78} (1997) 3129.

\bibitem{nepomnyashchy}  GOLOVIN A. A., DAVIS S. H., NEPOMNYASHCHY A. A.,
{\sl Physica D}, {\bf 122} (1998)202.

\bibitem{bodenschatz}  RAGNARSSON R., FORD J.L., SANTANGELO C. D. and
BODENSCHATZ E., {\sl Phys. Rev. Lett.}, {\bf 76} (1996) 3456.

\bibitem{Chevallard2000}  CHEVALLARD C., CLERC M., COULLET P.and GILLI J.M.,
{\sl Eur. Phys. J. E.}, {\bf 1} (2000) 179.

\bibitem{limat}  LIMAT L., {\sl Europhys. lett.}, {\bf 44} (1998) 205.

\bibitem{kuramoto}  KURAMOTO Y. , TSUZUKI T., {\sl Prog. Theor. Phys.}, {\bf
55} (1976) 356; SIVASHINSKY G.I., {\sl Acta Astronautica}, {\bf 4} (1977)
1177.

\bibitem{CocoMar}  CHEVALLARD C. and CLERC M., in preparation.

\bibitem{helfrich}  HELFRICH W., {\sl Phys. Rev. Lett.}, {\bf 21} (1968)
1518.

\bibitem{gilli}  BULAEVSKII L. N. and GINZBURG V. L., {\sl Zh. Eksp. Teor.
Fiz.}, {\bf 45} (1963) 772. [{\sl Sov. Phys. JETP}, {\bf 18} (1964) 530];
COULLET P., LEGA J., HOUCHMANZADEH B. and LAJZEROWICZ J., {\sl Phys. Rev.
Lett.}, {\bf 65} (1990) 1352.

\bibitem{cahn}  CAHN J. W., HILLIARD J. E., {\sl J. Chem. Phys.}, {\bf 28}
(1958) 258.

\bibitem{kawasaki}  KAWASAKI T., MUNAKATA T., {\scriptsize Prog. Theor. Phys.
}, {\bf 74} (1988) 656; KAWASAKI T., OHTA T., {\sl Physica A}, {\bf 11}
(1982) 573.

\bibitem{Bates}  ALIKAKOS N., BATES P. W., FUSCO G., {\sl J. Diff. Eqs.},
{\bf 90} (1990) 81; BATES P. W., XUN J. P., {\sl J. Diff. Eqs.}, {\bf 111}
(1990) 421.

\bibitem{Elliot}  ELLIOT C. M., FRENCH D., {\sl IMA Journal Appl. Math.},
{\bf 38} (1987) 97; EILBECK J.C., FURTER J.E., GRINFELD M., {\sl Phys. lett.
A}, {\bf 135} (1989) 272.


\end{references}
\end{document}